\documentclass{appolb}

\usepackage{amsmath,amssymb,bbm,color}
\usepackage{amsbsy}
\usepackage{euscript}
\usepackage{graphicx}
\usepackage{hyperref}
\usepackage{cite}
\usepackage{enumerate}

\newcommand{\order}[1]{${\cal O}(#1)$}

\begin{document}
\eqsec  
\title{%
\vspace{-4.0cm}
\begin{flushright}
		{\small {\bf IFJPAN-IV-2019-16}}   \\
\end{flushright}
\vspace{0.5cm}
\bf Theory challenges at future lepton colliders%
\thanks{Presented by S.J. at Matter To The Deepest
Recent Developments In Physics Of Fundamental Interactions
XLIII International Conference of Theoretical Physics.\\
This work is partly supported by
 the Polish National Science Center grant 2016/23/B/ST2/03927
 and the CERN FCC Design Study Programme.
}%
}
\author{Stanis\l{}aw Jadach and Maciej Skrzypek
\address{Institute of Nuclear Physics, Polish Academy of Sciences, 
         31-342 Krak\'ow, ul. Radzikowskiego 152, Poland}
}
\maketitle
\begin{abstract}
High energy, high luminosity, future lepton colliders, circular or linear,
may possibly give us hint about fundamental laws of Nature governing
at very short distances and very short time intervals, 
the same which have brought our Universe to live.
Currently considered projects are on one hand
linear electron-positron colliders, 
which offer higher energy and lower beam intensities
and on the other hand circular electron-positron colliders,
limited in energy but offering tremendous interaction rates.
On the far future horizon, muon circular colliders are the only viable
projects which can explore $>10$TeV teritory of the lepton colliders.
Experiments in all these future colliders will require theoretical calculations,
mainly of Standard Model processes (including QED), 
at the precision level one or even two orders better than available today.
After briefly characterization of theory puzzles in the fundamental
interactions we shall overview main challenges in the precision
calculations of the Standard Model effects, 
which have to be removed from data,
in order to reveal traces of new unexpected phenomena.
\end{abstract}


\section{Introduction}
High energy colliders considered for the future construction
and exploitation  would collide
hadron (proton) beams or lepton (electron, muon) beams.
What are presently main proposals for the future lepton colliders worldwide?
The leading candidates are: 
circular $e^+e^-$ collider FCC-ee~\cite{Abada:2019lih,Abada:2019zxq} 
in CERN delivering huge crop of events from 150 inverse attobarns (150$ab^{-1}$)
at 91GeV to 1.5$ab^{-1}$ at 365GeV~\cite{Blondel:2019yqr},
compact linear $e^+e^-$ collider CLIC in CERN
hopefully providing 1$ab^{-1}$ at 380GeV and up to 2.5$ab^{-1}$ at 1.5TeV,
international linear collider ILC in Japan which may deliver 2$ab^{-1}$
at 250GeV and 4$ab^{-1}$ at 500GeV,
finally another circular $e^+e^-$ collider CEPC in China
which would get 16$ab^{-1}$ at 91GeV and 6$ab^{-1}$ 
at 240GeV~\cite{BiscariGranada:2019}. 
Muon circular coliders with $\mu^+\mu^-$ beams will remain
the only option for another lepton collider at energies $\sim 15$TeV,
thanks to much weaker energy loss due to bremsstrahlung.
The most precise measurement of cross sections, asymmetries,
extremely rare decays would come from circular collider FCC-ee.
It will be able to provide $5\times 10^{12}$ Z boson decays,
$10^8$ WW events, $10^6$ HZ events and $10^6$ top quark pairs~\cite{Blondel:2019yqr}.
This is several orders of magnitude more than in previous
similar experiments (as compared to numbers of $Z$ and $WW$ at LEP),
reducing experimental errors of cross sections, asymmetries, masses,
decay rates by factor $10-100$.
As we shall see in the following,
it will be an enormous challenge to perform theoretical calculations
for these observables, within the Standard Model (SM) and beyond,
in order to match the above anticipated experimental precision.
In this short note we shall be able to overview only the main problems
of the above theory challenges.

\section{Puzzles of the fundamental physics}
Succesful experimental verification of 
the SM of the electroweak and strong interactions, 
the absence of direct signs of New Physics at multi-TeV experiments
at proton-proton collider LHC,
discovery of striking new properties of neutrinos,
and a wealth of new observations in astrophysics,
result in a number of burning questions on the nature of the fundamental
laws governing our Universe~\cite{ManganoCDR:2019}.
Theorists are deeply woried that the Nature has different opinion
about the ``naturalness'' then we do:
Higgs dynamics at the scale of the electroweak symmetry breaking
requires to be protected by the very ``un-natural'' fine tuning
of the dynamics at higher energies (shorter distance)
This is also called ``hierarchy problem''.
Moreover, we have no clue why do we have three families quark and leptons
and there is no systematic theoretical explanation of their masses and mixings.
Recent discovery of the neutrino masses and mixings add to the confusion.
Meeting point of gravity and quantum mechanics is still not understood.
According to accumulated knowledge, what we see as today's Universe was shaped
to a large extent at the end of the ``inflation era''
in the early stage of the ``Big Bang'',
but we do not know the origin and the details of the inflation,
except that it has to be closely related to Higgs dynamics~\cite{ServantGranada:2019}.
The mechanism of producing striking matter-antimatter asymmetry
in the present Universe still beggs explanation.
Better knowledge of the Higgs potential parameters 
would be valuable for the inflation modelling.
In particular $\sim 1\%$ measurement of the triple Higgs coupling
in the collider experiment would be of great interest for astrophysics.
The existence of the abundant Dark Matter everywhere in the Universe,
interacting gravitationaly with the ordinary matter, is another great puzzle.
Last but not least, the hypothetic Dark Energy speeding up expansion
of the Universe remains completely unexplained.

There is presently no satisfactory theory candidate, which could explain
the above puzzling phenomena and new hints from experiments are badly needed.
Experiments in high energy colliders are the most promissing source of such a hint.
At high energy colliders one may possibly see new particles and/or discover new
interactions of the known particles --
in particular decays of known unstable particles into forbidden final states
could provide valuable hint.
Very precise measurements of the properties of the known particles may depart
from the SM predictions, signaling new types of forces,
or existence of unknown much heavier particles.

\begin{figure}[!ht]
\centerline{%
\includegraphics[width=12.5cm]{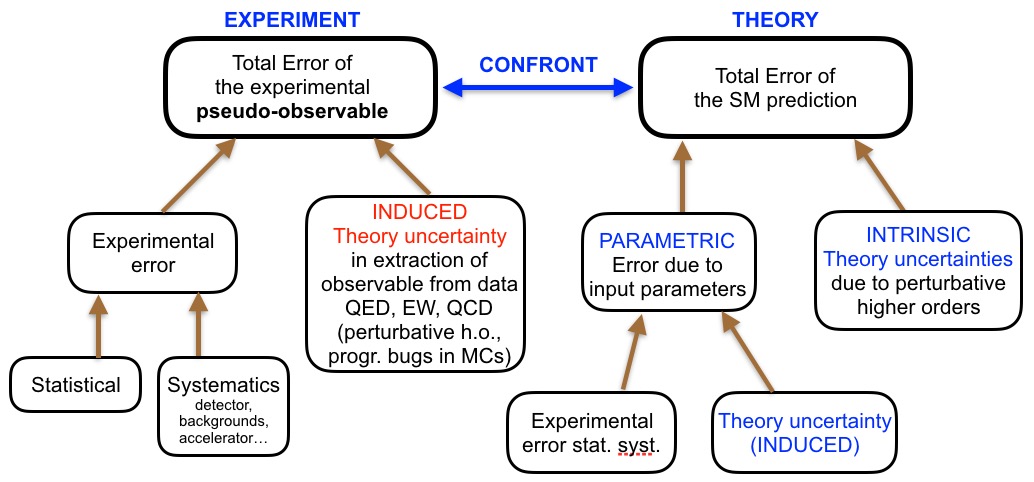}}
\caption{Scheme of the error propagation in the SM calculations.}
\label{fig:SMerrors}
\end{figure}

\section{Precision measurements of the electroweak observables}

The role of the SM theoretical predictions
for the future lepton collider measurements will be different
from the past role. 
In the past testing and verifyfying SM was the main aim.
In the future experiments SM will be assumed to be correct,
while searching for the deviation of the SM predictions from data 
as a sign of New Physics will be the main objective.
SM will be the tool and not the aim.
Perturbative calculations within the SM are comonly organised
in such a way that internal (Lagrangian level) SM parameters%
\footnote{Most important are three gauge coupling constants, 
   EW symmetry breaking scale
   and mass of the heaviest t-quark.
   Fermion masses and mixings also have to be added.}
are determined using limited number of the {\em SM input parameters}.
Typicaly they are observables known most precisely, for instance
$Z$ mass, the electromagnetic coupling $\alpha_{QED}(M_Z)$,
Fermi constant $G_\mu$ and the strong coupling constant $\alpha_S$.
Mass of the t-quark plays important role as input parameter.
All other observables, cross sections, asymmetries, 
masses of $W$ boson and Higgs boson $H$,
width and decay rates of $Z$, $W$ and $H$
can be calculated perturbatively,
in principle with an arbitrary precision
and will be confronted with the experiment~\cite{Blondel:2019vdq}.

SM input parameters are known with a certain experimental error,
which propagates to all SM predictions, and are called {\em parametric errors}.
The additional uncertainty of the SM predictions
due to technical uncertainties of the perturbative calculations
(uncalculated higher orders, numerical problems)
are commonly referred to as {\em intrisic errors}.
The map of errors in the SM calculations is shown schematically
in Fig.~\ref{fig:SMerrors}.

Alternatively, all observabels can be treated the same way in the global
fit of all observables to the SM, 
without any of them playing a privileged role of SM input parameters.

For more discussion the reader may consult 
Refs.~\cite{Strategy:2019vxc,Blondel:2019vdq,Blondel:2019qlh,Freitas:2019bre}.

\section{High precision SM calculations -- the role of resummation}
First order SM perturbative corrections split nicely into sum of three groups: 
``photonic'' QED real and virtual (loop) corrections,
pure EW loop corrections with heavy particles and QCD corrections.
Beyond the 1-st order this split gets fuzzy due to mandatory use of
soft and collinear resummation of QED higher orders,
renormalization group use in QCD
and the presence of QCD insertions in the EW multiloops.
Nevertheless, it is usefull to maintain it in practice, as far as it is possible.

Let us briefly chracterize genuine EW loop corrections,
omitting QCD component from our discussion,
while on the QED class we shall elaborate in a more
detail in the following sections.

EW loop corrections are relatively small, of order $\sim 1\%$,
as compared to QED effects which are of $\sim 10\%$ order.
The \order{\alpha^1} pure EW loop corrections 
to $e^+e^-\to f\bar{f}, W^+W^-$ processes
were completed at the start of LEP experiments in 1989 by several groups, 
but only two calculations embeded in the codes 
DIZET~\cite{Bardin:1989tq} (ZFITTER~\cite{Bardin:1999yd}) 
and TOPAZ0 \cite{Montagna:1998kp}
were used in the LEP data analysis \cite{ALEPH:2005ab,Schael:2013ita}.
It took another decade to complete most of \order{\alpha^2}
corrections to $e^+e^-\to f\bar{f}$ process \cite{Awramik:2006uz,Hollik:2006ma},
but only recently missing bosonic 2-loop corrections 
to $e^+e^-\to f\bar{f}$ process 
were calculated~\cite{Dubovyk:2018rlg, Dubovyk:2019szj}.
Generally, pure EW corrections are harder to calculate than QCD or QED corrections,
because of their multi-scale character -- 
with masses of gauge bosons, Higgs boson and all fermions
spanning over the entire range from 0.5MeV to 175GeV,
hence one cannot profit from smallness of some of them like in QCD or QED.
Consequently, the phase space of loop integrals 
has to be calculated without any approximations.
Sontaneous symmetry breaking and more complicated gauge group add to the problems.
In most complicated cases, like bosonic 2-loops analytical calculations,
analytical integrations are not feasible --
only numerical integration methods 
are able to cope~\cite{Dubovyk:2018rlg, Dubovyk:2019szj}.

In view of the 0.003\% precision for some observables near Z peak in FCC-ee
complete calculations of the \order{\alpha^3} EW corrections,
including 3-loop amplitudes will be needed.
It looks that again only numerical integration methods may work
for the integrations over 3-loop virtual phase space~\cite{Blondel:2019vdq}.
Such calculations will be rather slow and will have limited numerical precision,
but it is argued that 2-digit precision is good enough.

As soon as the complete \order{\alpha^1} 
EW corrections to $e^+e^-\to f\bar{f}$ process have been available, 
it was quite clear that their practical usefullness for the analysis
of the LEP data near Z peak is severly limited 
due to numericaly huge size of the pure QED component.
Even the upgrade of QED part to \order{\alpha^2} was not sufficient --
only after the inclusion of soft photon resummation, 
the desired theoretical precision $\sim 0.1\%$ 
for this process near the Z resonance was attained~\cite{Jadach:1992aa}
(similarly  $\sim 0.5\%$ precision for the $W$-pair production process).
In other words, the conservative order-by-order perturbative approach
does not work in practice --
one has to go to much higher perturbative orders 
for QED and QCD subclass of correction
(even to infinite order for soft photons)
than for the genuine pure EW parts.
The immediate question is therefore, how to disentangle in a systematic way
the QED part  and the so-called pure EW corrections,
performing IR cancellations within the soft photon resummation.
This question is especially intriguing  beyond the \order{\alpha^1},
where in a single diagram both photonic QED part
and genuine EW parts may show up simultaneously.

The solution of this problem is described and implemented in the so-called CEEX scheme,
see Refs.~\cite{Jadach:1999vf,Jadach:2000ir},
see also chapter C in Ref.~\cite{Blondel:2018mad}.
In the KKMC program~\cite{Jadach:1999vf,Jadach:2000ir},
the pure EW (non-soft) corrections are complete to \order{\alpha^1}, 
QED corrections are complete to \order{\alpha^2}
and soft photon corrections are resummed to infinite order.
The same scheme will work at higher orders, for instance for
genuine \order{\alpha^2} or \order{\alpha^3} EW corrections
combined with sufficiently higher order complete QED non-soft corrections
and soft photon resummation.
Moreover, this technique is implementable in the form of the Monte Carlo event 
generators
which can provide SM predictions  for arbitrary experimental cut-offs 
and/or detector efficiencies.
It is formulated using spin amlitudes,
so it works perfectly well for polarized initial and final particles.
It can also accommodate resummation 
of the coherent initial-final state interferences~\cite{Jadach:2018lwm}
for narrow neutral/charged resonances and can also deal with multiple photon emission
from unstable charged particles before they decay~\cite{Jadach:2019wol}.

However, in order to profit from the above technique one has to
calculate multi-loop SM corrections with QED component in a special way.
For instance in the \order{\alpha^2} SM calculations for $e^+e^-\to l^+l^-$ process
{\em one should not use Bloch-Nordsieck technique} of cancelling IR singularities
by means of adding (i) IR-divergent 2-loop contribution with one virtual photon line and
(ii) fully exclusive one-loop EW amplitudes
for to $e^+e^-\to l^+l^-\gamma$ subprocesses (without IR singularity inside the virtual part)%
\footnote{As it is done for instance in ref.~\cite{Khiem:2014dka}.
}.
In the CEEX scheme~\cite{Jadach:1999vf,Jadach:2000ir},
the well known IR component is {\em subtracted} from both above corrections,
because the IR cancellation is executed indepentently
within the soft photon resummation part of the calculation.

\section{QED challenges at FCC-ee precision}

\begin{table}
\centering\small
\begin{tabular}{|c|c|c|c|c|c|}
\hline
 Observable & from & Present (LEP) & FCC stat. & FCC syst& 
$\frac{\rm Now}{\rm FCC}$\\
\hline
$M_Z$ [MeV]    & $Z$ linesh.
               & $91187.5\pm 2.1\{0.3\}$ & $0.005$ & $0.1$ 
               & 3 \\
$\Gamma_Z$ [MeV] & $Z$ linesh.
               &  $2495.2\pm 2.1\{0.2\}$ & $0.008$ & $0.1$ 
               & 2\\
$R^Z_l=\Gamma_h/\Gamma_l$ & $\sigma(M_Z)$
               & $20.767\pm 0.025\{0.012\}$& $ 6\cdot 10^{-5}$&$ 1\cdot 10^{-3}$
               & 12\\
$\sigma^0_{\rm had}$[nb] & $\sigma^0_{\rm had}$ 
               & $41.541 \pm 0.037\{0.025 \}$
               & $ 0.1\cdot 10^{-3}$ & $ 4\cdot 10^{-3}$ 
               & 6\\
$N_\nu$        & $\sigma(M_Z)$
               & $2.984\pm 0.008\{0.006\} $  &$ 5\cdot 10^{-6}$ &$1 \cdot 
10^{-3}$
               & 6\\
$N_\nu$        & $Z\gamma$
               & $2.69\pm0.15\{0.06\} $ & $ 0.8 \cdot 10^{-3}$& $<10^{-3}$ 
               & 60\\
$\sin^2\theta_W^{eff}\!\times 10^{5}$ 
               & $A_{FB}^{lept.}$
               & $23099\pm 53\{28\}$ & $ 0.3$ &  $ 0.5$ 
               & 55\\
$\sin^2\theta_W^{eff}\!\times 10^{5}$ 
               &$\langle{\cal P}_\tau\rangle$,$A_{\rm 
FB}^{pol,\tau}\!\!$
               & $23159\pm 41\{12\}$ & $0.6$ & $<0.6$ 
               & 20 \\
$M_W$ [MeV]    & ADLO
               & $80376\pm 33\{6\}$  &   0.5    &  0.3   
               & 12 \\
 {\small $A_{FB,\mu}^{M_Z\pm 3.5 {\rm GeV}}$}
 & $\frac{d\sigma}{d\cos\theta}$
               & $\pm 0.020\{0.001\}$
               & $1.0\cdot 10^{-5}$  & $0.3\cdot 10^{-5}$  
               & 100 \\
\hline
\end{tabular}
\caption{\sf
 Table of electroweak observables
 most sensitive to QED effects from ref.~\cite{Jadach:2019bye}.
 LEP experimental errors (3-rd column) are accompanied
 in the braces $\{...\}$ by the induced QED uncertainties.
 FCC-ee experimental systematic errors in 4-th column are
 from FCC-ee CDR~\cite{Mangano:2018mur}
 except $\tau$ polarisation~\cite{2019:cern-talk-tenchini}.
 The improvement factor in QED theoretical calculations
 needed to equalize with experimental precision of FCC-ee measurements
 is shown in the last column.
}
\label{tab:QEDchal}
\end{table}

Trivial but numericaly sizable pure QED effects were removed
from all LEP data (observables) like $Z$ and $W$ masses and widths,
cross sections, asymmetries, decay widths.
However, this was resulting in the induced QED uncertaity
in the experimental errors of these observables.
These QED uncertaities were usually at least factor three smaller
than the genuine statistical and systematic experimental errors.
This can be seen in the 3-rd column of Table~\ref{tab:QEDchal}
taken form Ref.~\cite{Jadach:2019bye}.
Next columns in the table show the enormous progress, up to factor 100,
in experimental errors to be attained in FCC-ee experiments.
Obviously, the precision of the theoretical QED calculations
have to progress at least to the level of the FCC-ee experiments.
The corresponding minimal improvement factor of the QED calculations 
is shown in the last column of Tab.~\ref{tab:QEDchal}.
In fact, this factor has to be about three times better,
if we want to be in the same comfortable situation as in LEP.
The same information as in Tab.~\ref{tab:QEDchal}
is also visualised in Fig.~\ref{fig:SMerrors}.

\begin{figure}[!h]
\centerline{%
\includegraphics[width=130mm]{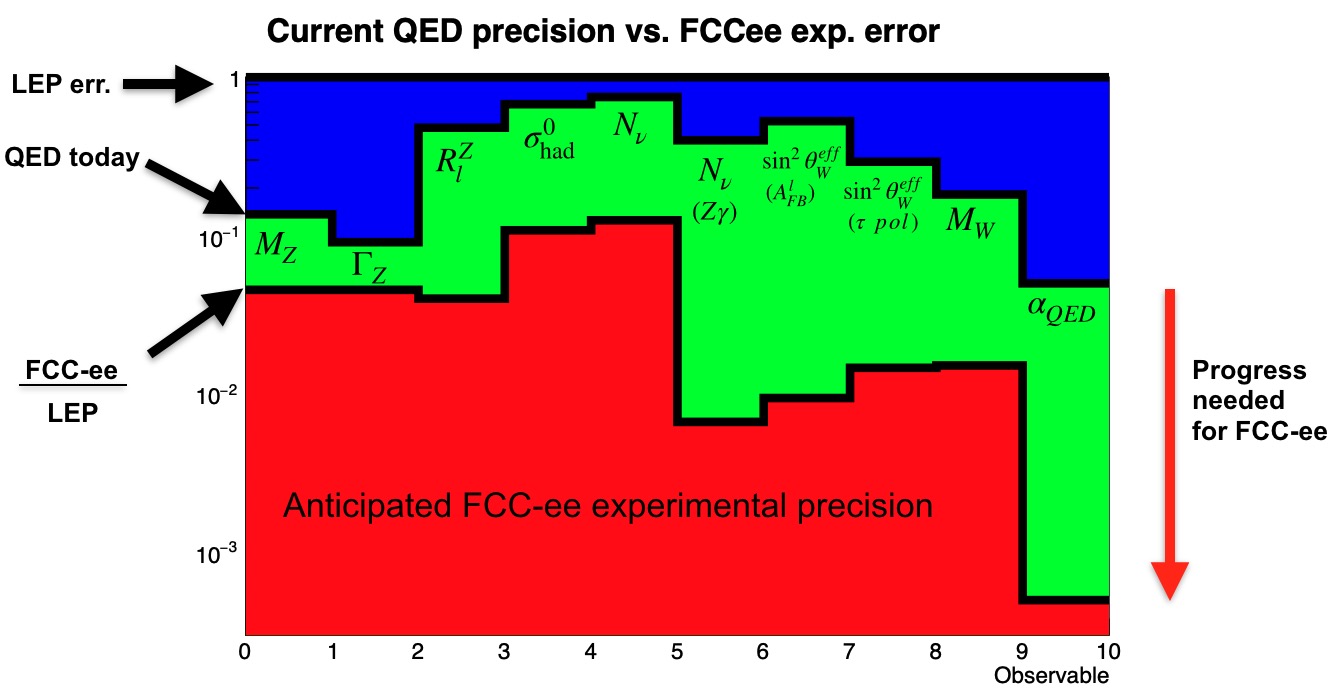}}
\caption{QED challenges at FCC-ee of Tab.~\ref{tab:QEDchal}
 in a graphical form. }
\label{fig:QEDchal}
\end{figure}

Before we can discuss whether the desired improvement of the QED
calculations shown in Table~\ref{tab:QEDchal} 
and  Fig.~\ref{fig:SMerrors} is feasible,
one has to answer even more basic and highly non-trivial question
whether the methodology of removing QED effects from the listed
observables, which was used at LEP data analysis, 
will still work at the tremendous precision of FCC-ee?
This question will be briefly elaborated in the next section --
for more detailed disctussion we refer the reader 
to Ref.~\cite{Jadach:2019bye}.

The question whether improvement factors of QED calculations 
in Tab.~\ref{tab:QEDchal} are achievable is discussed
at length in Ref.~\cite{Jadach:2019bye}.
Here, let us only summarize briefly on that in the following.
The important point is that contrary to LEP data analysis,
where semi-analytical programs like ZFITTER or TOAPZ0 have
played major role, at FCC-ee only Monte Carlo
calculations will be able to provide cut-off dependent SM predictions with 
sufficient precision%
\footnote{Perhaps with the only exception of total hadronic cross section.}.
Moreover fitting data to full SM prediction 
or to ``effective SM parametrization'' of data in form
of EW pseudo-observables (see next section)
will be also done using MC programs.

Just to give a few examples, near the $Z$ peak improvements in precise
measurement of the hadronic total cross section providing experimental
$~\delta M_Z,\delta \Gamma_Z \le 0.1\;$MeV,
will require QED to be reduced to $\leq 0.03\;$MeV i.e. by factor 10.
Better modelling of light fermion production and  the inclusion of 
${\cal O}(\alpha^2 L_e^0, \alpha^3L_e^2,\alpha^4L_e^4 ),\; L_e=\ln(s/m_e^2)$
initial state QED corrections will be mandatory.
Data analysis for final leptonic states near $Z$ resonence will
be more demanding.
In the MC programs of the KKMC class with CEEX matrix element
at least the inclusion of \order{\alpha^2 L_e} penta-boxes 
and of \order{\alpha^3L_e^3} photonic corrections will be necessary.
Provisions for SM parameter fitting  and extracting EWPOs from data
will have to be included in the MC programs.
Measurement of charge asymmetry with the experimental error
$\delta A^\mu_{FB}(M_Z) \simeq 1\cdot 10^{-5}$,
leading to $\delta\sin^2\theta_W^{eff} \simeq 0.5\cdot 10^{-5}$
will require factor 50-150 improvement in the control of QED effects.
Such improvements are particularily urgent for the Bhabha process.
Simmilarily, the anticipated experimental error
$\delta\sin^2\theta_W^{eff} \simeq 0.6\cdot 10^{-5}$
from spin asymmetry measurements in tau pair production and decay
at FCC-ee will require factor 20-60 better understanding of QED effects.
As seen in Tab.~3 in Ref.~\cite{Blondel:2019qlh} the precision
of the QED coupling constant $\alpha_{QED}(M_Z)$, as an input in
the SM calculations, 
is critical for precision of all SM predictions~\cite{Blondel:2019qlh}.
In Ref.\cite{Janot:2015gjr} it was proposed to extract $\alpha_{QED}(M_Z)$
from the measurment of $A_{FB}(M_Z\pm 3.5$GeV) with precision $3\cdot 10^{-5}$, 
that is factor 200 more precisely than at LEP.
However, the QED Initial-Final state interference IFI is here the main obstacle!
Wile IFI cancels partly in the difference of  $A_{FB}(M_Z\pm 3.5$GeV),
the ~1\% effect remains in $A_{FB}(M_Z\pm 3.5$GeV).
Can one control IFI in the charge asymmetry near Z resonance
with the precision $3\cdot 10^{-5}$?
In  ref.~\cite{Jadach:2018lwm} it was shown, using KKMC 
and new KKfoam programs, that one may reach precision $10^{-4}$.
More effort is needed to get another improvement factor of 10.

The precision determination of the luminosity using low angle Bhabha process
at FCC-ee will be again limited by the knowledge of higher order QED effects
and hadronic contributions to vacuum polarization (VP) correction.
In Ref.~\cite{Jadach:2018jjo} and Chapter B of Ref.~\cite{Blondel:2019vdq}
it was shown that $10^{-4}$ precision of theoretical calculation
of the low angle Bhabha for FCC-ee luminometer is feasible.
This will allow to reduce error of the invisible $Z$ decay rate measured
in terms of the ``number of neutrinos'' $N_\nu$ from present 
$\delta N_\nu = \pm 0.006$ down to $\delta N_\nu = \pm 0.001$.
Similar precision of $N_\nu$, also limited by the QED effects, 
will be achievable using process $e^+e^-\to X \gamma,\; X\to\; invis.$
\cite{ifjpan-iv-2019xx}.

New more precise calculation of the $e^+e^-\to W^+W^-$ is needed for the FCC-ee
measurements of $W$ mass and couplings.
The 0.5MeV precision of $W$ mass from the threshold cross section 
and the mass distributions in the final state will require 
clever resummation of the QED effects using QED resummation 
techniques~\cite{Jadach:2019wol},
Effective Field Theory~\cite{Actis:2008rb,Blondel:2019vdq}
and new higher order EW claculations beyond the \order{\alpha^1}~\cite{Denner:2005fg}.
Precise measurement of the $WW$ cross sections (distributions)
and W mass ($\sim 0.5$MeV) will require:
(i) \order{\alpha^2} calculation of EW corrections for double-resonant (on-shell) 
  -- non-trivial but feasible, to be done,
(ii) \order{\alpha^1} calculation for single-resonant component 
     -- (partly done in Ref.~\cite{Denner:2005fg}),
(iii) tree-level for non-resonant part (available),
(iv) and the consistent scheme of combining all that within the Monte Carlo event generator!
QED component will be again most sizeable%
\footnote{As in the LEP era calculations 
 of RACOONWW~\cite{Denner:2000bj} and YFSWW+KORALW~\cite{Jadach:2001mp}.}
and equally important as pure EW corrections.

\section{The need of new ideas for EW pseudo-observables}

In the LEP era data analysis based on ref.~\cite{Bardin:1999gt}
and summarized in Refs.~\cite{ALEPH:2005ab,Schael:2013ita}
there were two types of observables,
realistic observables (ROs), i.e.
cross sections and distributions for well defined
experimental cut-offs 
(after removing detector inefficiencies using Monte Carlo)
and EW pseudo-observables (EWPOs),
in which QED effects were removed (deconvoluted).
The simplest example of EWPO is hadronic (or total) cross section
exactly at the mass of $Z$, $\sigma^0_{had}$.
It was obtained in LEP in such a way that experimental cross section
at seven energy points was fitted with the following formula
\begin{equation}
\sigma_{had.}(s) = \int_0^1 dz\; \sigma^{Born}_{had.}(zs)\; \rho_{QED}(z),
\end{equation}
where $\sigma^{Born}(s)$ comes from analytical formula
in eq.~3.8 in  Ref.~\cite{ALEPH:2005ab}.
Mass and width of $Z$ and couplings of $Z$ to electron and final quarks
are also obtained from the same fit.
The effective radiator function $\rho_{QED}(z)$
represents perturbative QED result for the initial state multiphoton radiation.
Finally, hadronic cross section $\sigma^0_{had}$ is obtained
from analytical formula $\sigma^0_{had}\equiv\sigma^{Born}_{had.}(M_Z^2)$
inserting into it all parameters from the fit to data.
Leptonic cross section $\sigma^0_l$ is obtained the same way.

Similarly, the charge asymmetry for lepton pair production process 
$e^+e^-\to l^+l^-$ is obtained using another convolution formula
\begin{equation}
\frac{d\sigma^\mu}{d\cos\theta^*}(s,\theta^*) 
= {\rm\bf CONV}\Big\{ \frac{d\sigma^{Born}_\mu(s)}{d\cos\theta}, \rho_{QED} \Big\},
\end{equation}
where $\theta^*$ is some experimentally well defined
effective angle of outgoing leptons
(they are not back to back due to photon emission).
The meaning of the convolution ${\rm\bf CONV}$
and the definition of the analytical formula for the effective Born distribution
can be found in Ref.~\cite{ALEPH:2005ab}.
The value of the pseudo-observable charge asymmetry $A_{FB}^l$
does not correspond directly to asymmetry of some well defined 
experimental alngular distribution, 
but results from the following analytical formula
\begin{equation}
 A_{FB}^l =\frac{3}{4} {\cal A}_e {\cal A}_l,\quad
 {\cal A}_f = \frac{g_{Lf}^2-g_{Rf}^2}{g_{Lf}^2+g_{Rf}^2},
\end{equation}
where again the values of $Z$ couplings $g_{Lf}$ and $g_{Rf}$ to fermion $f$
are determined in the fit of 
$\frac{d\sigma^\mu}{d\cos\theta^*}(s,\theta^*) $ to data.
The effective EW mixing angle is obtained also from the simple
analytical formula
\[
\frac{g_{Vf}}{g_{Af}}= 1-\frac{2Q_f}{T^3_f}\sin^2\theta^f_{eff}
\]
using fitted values of $Z$ couplings.

In a similar way, that is using simple analytical formula with fitted
$Z$ couplings, mass and width inside, 
all nine EWPOs listed in Tab.~2.4 in Ref.\cite{ALEPH:2005ab}
$m_Z,\Gamma_Z,\sigma^0_{had},R^0_e,R^0_\mu,R^0_\tau,A_{FB}^{0,e},A_{FB}^{0,\mu},A_{FB}^{0,\tau}$
were obtained.
The fundamental role of these EWPOs was to encapsulate in a compact way experimental data,
such that SM predictions including \order{\alpha} EW corrections were confronted
with cut-off independent EWPOs with the removed QED effects,
instead of cut-off dependent realistic data including QED effects.

\begin{figure}[!h]
\centerline{%
\includegraphics[width=120mm]{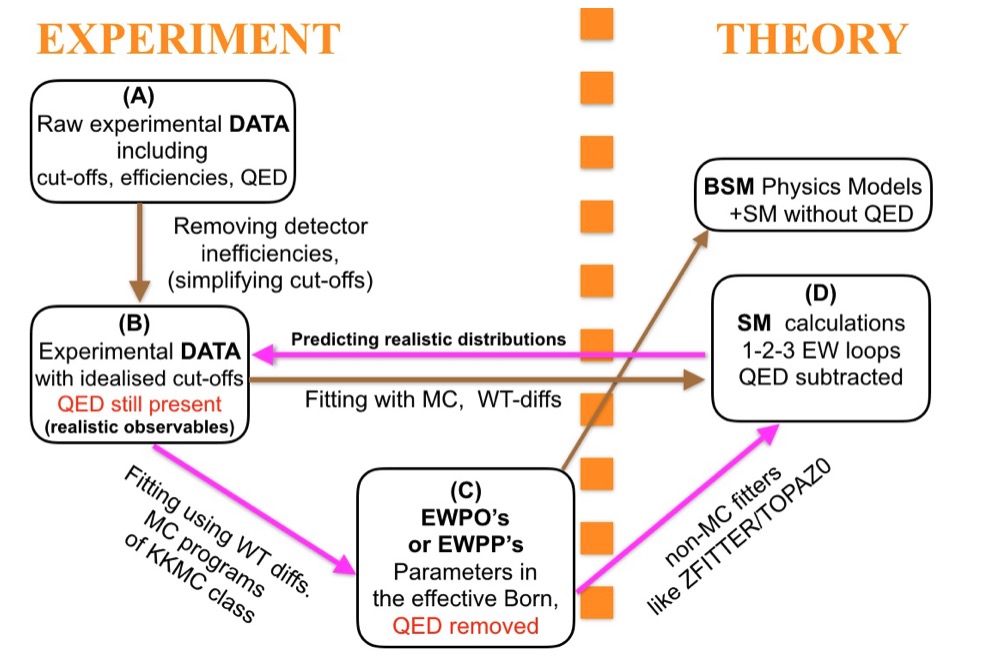}}
\caption{ Scheme of construction of EWPOs at FCC-ee.
 Main difference with LEP is Monte Carlo use 
 in steps (B)$\to$(C) and (B)$\to$(D) instead of progs like ZFITTER/TOPAZ0
}
\label{fig:EWPO}
\end{figure}

Of course, the use of EWPOs at LEP was dangerous, because the convolution formulas
were including simplified version of QED calculation 
(without initial-final state interference and with fully inclusive treatment
of the final state radiation).
The use of the effective Born with effective $Z$ couplings could also be
incompatible, at a certain precision level, with the presence 
of the \order{\alpha^1} EW corrections in the data 
(additional angular dependence from EW boxes).
In Refs.~\cite{ALEPH:2005ab,Bardin:1999gt} it was proven that
at the LEP data precision such dangers were avoided,
by means of comparing realistic data with the predictions of the SM, 
in which internal parameters were previously fit to EWPOs.
Such a ``circular cross-check'' is illustrated in Fig.~\ref{fig:EWPO}.

It is quite likely that the above LEP construction of EWPOs will not pass
the above circular cross-check test 
due to much smaler errors of FCC-ee experimental data.
How to upgrade the definitions of EWPOs,
such that they works at the FCC-ee precision level?
Answering this question requires dedicated study.
Most likely two elements will have to be modified.
In the transition from realistic data in step (B)
to new EWPOs in step (C) in Fig.~\ref{fig:EWPO}
semianalytical codes like ZFITTER or TOPAZ0 will have to be replaced
by the Monte Carlo programs of the KKMC class, or even more sophisticated ones.
Most likely, the effective Born-like formula for spin amplitudes used to parametrize
data in the (B)$\to$(C) transition will have to include 
more of genuine \order{\alpha^1} EW corrections, 
removing them from the data in the form of new EWPOs,
in the same way as trivial pure QED effects.

\section{Summary and outlook}

We cannot get to better understanding of fundamental laws on Nature
without answering lot of big intriguing questions!
Unfortunately, there is no clear hint from theory where to look for the answers.
Hence one should explore all possible experimental fronts:
 highest possible energies,
 very weak and rare processes (neutrinos),
 astrophysics.

High precision measurements in the future electron-positron colliders will 
require
major effort in order to improve SM/QED predictions for FCC-ee observables by factor 10-200.
In particular precision of QED calculations of asymmetries 
near the $Z$ reasonance has to be improved by factor up to 200.
New algorithms of extracting EW pseudo-observables from experimental data 
has to be worked out and cross-checked.
The increased role of MC event generators
at all levels of data analysis and in comparisons with the theory is anticipated.


\end{document}